\documentclass[journal]{IEEEtran}

\usepackage[table]{xcolor}
\usepackage{colortbl}
\usepackage{xfp} % for floating-point comparisons
\usepackage{booktabs}

\definecolor{darkgreen}{RGB}{0,100,0}
\definecolor{yellowish}{RGB}{255,255,153}
\definecolor{redish}{RGB}{255,102,102}
\definecolor{revisionblue}{RGB}{0,0,180}

\newcommand{\AccCell}[1]{%
    \ifnum#1=100
        \cellcolor{darkgreen!60}\textbf{#1}%
    \else\ifnum#1>95
        \cellcolor{green!40}#1%
    \else\ifnum#1>84
        \cellcolor{yellowish}#1%
    \else
        \cellcolor{redish}#1%
    \fi\fi\fi
}

\newcommand{\PowCell}[1]{%
    \edef\val{#1}%
    \ifdim \val pt < 0.5pt
        \cellcolor{darkgreen!60}#1%
    \else\ifdim \val pt < 3pt
        \cellcolor{green!40}#1%
    \else\ifdim \val pt < 5pt
        \cellcolor{yellowish}#1%
    \else
        \cellcolor{redish}#1%
    \fi\fi\fi
}

\usepackage{pifont}
\usepackage{xcolor}
\usepackage{amsmath,amsfonts,amssymb}
\usepackage{array}
\usepackage[caption=false,font=normalsize,labelfont=sf,textfont=sf]{subfig}
\usepackage{textcomp}
\usepackage{stfloats}
\usepackage{url}
\usepackage{enumitem}
\usepackage{verbatim}
\usepackage{graphicx}
\usepackage{algorithm}
\usepackage{algpseudocode}

\usepackage{booktabs} % For formal tables
\usepackage{stackengine}
\usepackage{float}
 \usepackage[table,xcdraw]{xcolor}
\usepackage{tikz}
\usepackage{xcolor}
\usepackage{wasysym}
\usepackage{colortbl}
\usepackage{soul}
\usepackage{multirow}
\usepackage{ragged2e}
\usepackage{hyperref}
\usepackage{comment}
\usepackage{pifont} % For checkmark symbols
\usepackage{xcolor}  % For color support

\newcommand\encircle[1]{%
\tikz[baseline=(X.base)]
 \node (X) [draw, scale=0.75, shape=circle, inner sep=0, fill=black, text=white, minimum size=0em] {\strut #1};}

\title{\vspace{-0.8em}{LIMCA}: \ul{L}LM for Automating Analog \ul{I}n-\ul{M}emory \ul{C}omputing \ul{A}rchitecture Design Exploration \vspace{-0.3em}}
\author{Deepak Vungarala,~\IEEEmembership{Student Member,~IEEE,} 
        Md Hasibul Amin,~\IEEEmembership{Student Member,~IEEE,}\\
        Pietro Mercati,~\IEEEmembership{Member,~IEEE,}
        Arnob Ghosh,~\IEEEmembership{Member,~IEEE,}
        Arman Roohi,~\IEEEmembership{Senior Member,~IEEE,}\\
        David Z. Pan,~\IEEEmembership{Fellow,~IEEE,} 
        Ramtin Zand,~\IEEEmembership{Member,~IEEE,}
        and Shaahin~Angizi,~\IEEEmembership{Senior Member,~IEEE} 
        \thanks{\footnotesize  This work is supported in part by Synopsys Research Gift and the National Science Foundation (NSF) under grant no. 2228028, 2409697, and 2504839.}
        \vspace{-2.8em}
\IEEEcompsocitemizethanks{\IEEEcompsocthanksitem \footnotesize D. Vungarala, A. Ghosh, and S. Angizi are with the Department of Electrical and Computer Engineering, New Jersey Institute of Technology, NJ, USA.
E-mail: \{dv336,shaahin.angizi\}@njit.edu.
\IEEEcompsocthanksitem{P. Mercati is with the Intel Corporation, Hillsboro, OR, USA. E-mail: pietro.mercati@intel.com.}
\IEEEcompsocthanksitem M. H. Amin and R. Zand are with the Department of Computer Science and Engineering, University of South Carolina, SC, USA. E-mail: ramtin@cse.sc.edu.
\IEEEcompsocthanksitem{A. Roohi is with the Department of Electrical and Computer Engineering, University of Illinois Chicago, IL, USA. E-mail: aroohi@uic.edu.}
\IEEEcompsocthanksitem{D. Z. Pan is with the Department of Electrical and Computer Engineering, University of Texas at Austin, TX, USA. E-mail: dpan@ece.utexas.edu.}
}}

\vspace{-4.2em}%}

\begin{document}

% The paper headers
\markboth{IEEE Transactions on Computer-Aided Design of Integrated Circuits and Systems}%
{Shell \MakeLowercase{\textit{et al.}}: Bare Demo of IEEEtran.cls for IEEE Journals}

\maketitle

\begin{abstract}
In this paper, we introduce \texttt{LIMCA}, a novel \textit{fine-tuning-free} Large Language Model (LLM)-driven framework for automating the design and evaluation of In-Memory Computing (IMC) crossbar architectures. Unlike traditional approaches, where the manual, knowledge-intensive design process and the lack of high-quality circuit netlists have significantly constrained Design Space Exploration (DSE) and optimization to behavioral system-level tools, \texttt{LIMCA} employs a \textbf{No-Human-In-Loop (NHIL)} automated pipeline to generate and validate circuit netlists for SPICE simulations, eliminating manual intervention. \texttt{LIMCA} systematically explores the IMC design space by leveraging a structured dataset and LLM-based performance evaluation. Our experimental results on MNIST classification demonstrate that \texttt{LIMCA} successfully generates crossbar designs achieving \textbf{$\geq$96\% accuracy while maintaining a power consumption $\leq$3W}, making this the first work in LLM-assisted IMC DSE. Compared to existing frameworks, \texttt{LIMCA} provides an automated, scalable, and hardware-aware solution, reducing design exploration time while ensuring user-constrained performance trade-offs. Artifacts are open-sourced at \hyperlink{https://github.com/ACADLab/LIMCA}{https://github.com/ACADLab/LIMCA}
\end{abstract} \vspace{-0.25em}
\begin{IEEEkeywords}
In-memory computing (IMC), crossbar arrays, large language models (LLMs), design-space exploration.
\end{IEEEkeywords}\vspace{-1.5em}

\section{Introduction}\vspace{-0.5em}
\IEEEPARstart{T}{he} increasing complexity and computational demands of Deep Neural Networks (DNNs) have highlighted the limitations of traditional von Neumann architectures, particularly the memory wall bottlenecks in data movement between processing and memory units. To address this, Analog In-Memory Computing (IMC) crossbar architectures have emerged as a promising solution, offering the ability to perform computations directly within memory arrays, thereby significantly reducing data movement and energy consumption \cite{chen2018neurosim}. These architectures leverage the physical properties of resistive devices to store DNN weight parameters and perform matrix operations in the analog domain, enabling massive parallelization of DNN computations \cite{zhu2020mnsim}.
However, designing analog IMC systems presents unique challenges such as deep expertise in analog circuit design, an understanding of device physics, and careful consideration of various non-idealities such as parasitic effects, device variations, and noise \cite{amin2022xbar}. 

Traditional analog IMC design approaches \cite{chen2018neurosim,9458494} rely on manual optimization and iterative refinement, making it difficult to explore the vast design space of possible implementations as shown in Table \ref{comp}. Furthermore, the lack of standardized tools and methodologies for IMC design has hindered rapid prototyping and evaluation of novel architectures. Recently, Large Language Models (LLMs) have shown success in digital design automation and hardware code generation \cite{chang2023chipgpt,vungarala2024spicepilot,thakur2023verigen}. The capability of LLMs in generating von Neumann architecture-based Artificial Intelligence (AI) accelerators has been recently explored \cite{10323953,vungarala2024sa}. Nevertheless, their potential to generate promising non-von Neumann architectures, such as resistive IMC crossbar systems that allow for parallel and efficient vector-matrix multiplication, has remained unexplored. This gap stems from the knowledge intensive hardware design process and the scarcity of high-quality datasets and circuit catalogs due to proprietary data \cite{10323953,vungarala2024spicepilot}.

% \begin{table}[t] \vspace{-1em}
% \centering
% \textcolor{blue}{\caption{Comparison of LIMCA with State-of-the-Art Approaches}}
% \textcolor{blue}{
% \scalebox{0.7}{
% \begin{tabular}{@{}lcccc@{}}
% \toprule
% \textbf{Framework} & \textbf{Automation} & \textbf{Natural Language} & \textbf{Extrapolation} & \textbf{SPICE Fidelity} \\
% \midrule
% NeuroSim [1] & Manual & No & No & Behavioral \\
% MNSIM [2] & Manual & No & No & Behavioral \\
% AIHWKIT [3] & Manual & No & No & Behavioral \\
% CCCS [4] & Manual & No & No & SPICE-level \\
% IMAC-Sim [5] & Manual & No & No & SPICE-level \\
% Grid Search & Automated & No & No & N/A (lookup) \\
% Rule-Based & Automated & Limited & No & N/A (lookup) \\
% \textbf{LIMCA} & \textbf{Automated} & \textbf{Yes} & \textbf{Yes} & \textbf{SPICE-level} \\
% \bottomrule
% \end{tabular}}}
% \end{table}

\begin{table}[t]\vspace{-1em}
\centering
\caption{{State-of-the-art analog IMC simulation frameworks.}} \vspace{-0.7em}
\scalebox{0.70}{
\begin{tabular}{lccccc}
\hline
\multicolumn{1}{c}{Frameworks}             & Type    & \begin{tabular}[c]{@{}c@{}}Design Space\\  Exploration\end{tabular} & \begin{tabular}[c]{@{}c@{}}Support for\\ {PAA$^*$} cons.\end{tabular} & Language                         & \begin{tabular}[c]{@{}c@{}}Inference\\ Accuracy\end{tabular} \\ \hline
NeuroSim \cite{chen2018neurosim} & System  & Manual                                                              & No                                                                   & C++                              & estimate                                                     \\
MNSIM \cite{zhu2020mnsim}                  & System  & Manual                                                              & No                                                                   & Python                           & estimate                                                     \\
AIHWKIT \cite{9458494}                     & System  & Manual                                                              & No                                                                   & Python                           & estimate                                                     \\
CCCS \cite{zhang2020cccs}                  & Circuit & Manual                                                              & No                                                                   & \multicolumn{1}{l}{MATLAB-SPICE} & exact                                                        \\
IMAC-Sim \cite{Amin2023IMACSimAC}               & Circuit & Manual                                                              & No                                                                   & Python-SPICE                     & exact                                                        \\
DPE \cite{hu2016dot}                       & Circuit & Manual                                                              & No                                                                   & MATLAB                           & estimate                                                     \\
\rowcolor[HTML]{9AFF99} 
\texttt{LIMCA}                                      & Circuit & Automated                                                           & Yes                                                                  & Python-SPICE                     & {SPICE-char.}                                                        \\ \hline
\end{tabular} \vspace{-4em}}

\tiny $^*$ Power-Area-Accuracy optimization.
\label{comp}\vspace{-4.8em}
\end{table}

This work introduces the first LLM-driven automated design exploration framework for IMC crossbar architectures dubbed \texttt{LIMCA} to generate design under user defined Power, Area, and Accuracy (PAA) constraints as highlighted in Table \ref{comp}. The inherent challenges of IMC design including the need for precise circuit specifications, the consideration of complex analog behaviors, and the trade-offs among multiple competing objectives present both unique opportunities and challenges for LLM-based automation. Therefore, the core questions we seek to answer are the following-- 
\textit{(RQ-1)} \textit{Can we automate IMC design to address the growing shortage of specialized hardware in the semiconductor industry?}  
\textit{(RQ-2)} \textit{How can we deploy LLMs without cost-intensive fine-tuning? and can we effectively perform and evaluate Design Space Exploration (DSE)?}  
\textit{(RQ-3)} \textit{Can the major time consuming validation pipeline be fully automated?}
To answer these questions, this work presents the following contributions.
(1) A DSE framework, \texttt{LIMCA} with Automatic Design Generation that leverages LLMs to map domain knowledge and introduces a fine-tuning-free approach for generating user-constrained designs from the space and generating the design of their choice.  (2) \texttt{LIMCA} implements an automated validation strategy, eliminating human intervention, a No-Human-In-Loop (NHIL) approach ensuring efficient and scalable design evaluation. (3) An extensive open-source IMC dataset containing detailed variations of designs, along with PAA metrics for analog IMC.

{Beyond these contributions, \texttt{LIMCA} provides four unique capabilities that distinguish it from existing tools: $(i)$ \textit{Natural Language DSE Interface} users specify constraints in plain English rather than explicit parameter values; $(ii)$ \textit{Autonomous Design Navigation} unlike lookup-based approaches, \texttt{LIMCA} proposes design solutions for configurations not in the dataset by learning patterns from existing data points through In-Context Learning (ICL), then autonomously refining toward valid solutions via NHIL; $(iii)$ \textit{Constraint Relaxation Negotiation} when user constraints cannot be satisfied, \texttt{LIMCA} proposes alternative configurations with explicit trade-off explanations; and $(iv)$ \textit{Self-Correcting Pipeline} hold-out validation experiments demonstrate that \texttt{LIMCA} proposes solutions for 100\% of queries on unseen configurations.}

\vspace{-1.2em}

\section{Background, Challenges, and Motivations}\vspace{-0.3em}
\textbf{LLM-Driven Hardware Design.} 
LLMs have shown strong potential in automating Hardware Description Language (HDL) and High-Level Synthesis (HLS) code generation. Early works such as VeriGen \cite{thakur2023verigen} and ChatEDA \cite{wu2024chateda} improve RTL-to-GDSII flows. ChipGPT \cite{chang2023chipgpt} and Autochip \cite{thakur2023autochip} integrate LLMs for design generation and optimization, with Autochip refining Verilog through feedback.  
Dataset-driven frameworks have further advanced this direction. SA-DS \cite{vungarala2024sa} introduced an HLS dataset for DNN accelerators. Interactive approaches, such as Chip-Chat \cite{blocklove2023chip}, extend LLM applicability to DSE. Meanwhile, RTLLM \cite{lu2024rtllm}, GPT4AIGChip \cite{10323953}, and VerilogReader \cite{ma2024verilogreader} highlight efficiency improvements in design and coverage-driven code analysis. Domain-specific accelerator generation is addressed by only a few efforts. GPT4AIGChip \cite{10323953} targets HLS flows. On the analog domain, AnalogCoder \cite{lai2024analogcoder}, AnalogCoder-pro \cite{lai2025analogcoder}, SPICEPilot \cite{vungarala2024spicepilot}, and Masala-CHAI \cite{bhandari2025masalachailargescalespicenetlist} pioneer SPICE-level netlist generation through ICL and agentic frameworks.
%, while AmpAgent \cite{liu2024ampagent} focuses on multi-stage amplifier design and technology porting.  
Despite these advances, challenges remain. Most frameworks lack optimized prompting, tailored datasets, and model fine-tuning, while hallucination remains a persistent issue \cite{wu2024chateda,vungarala2024sa}. Addressing these limitations is key to fully realizing LLMs' potential in hardware and circuit design automation.

\begin{figure}[t]
\begin{center}\vspace{-2em}
\includegraphics[width=0.8\linewidth]{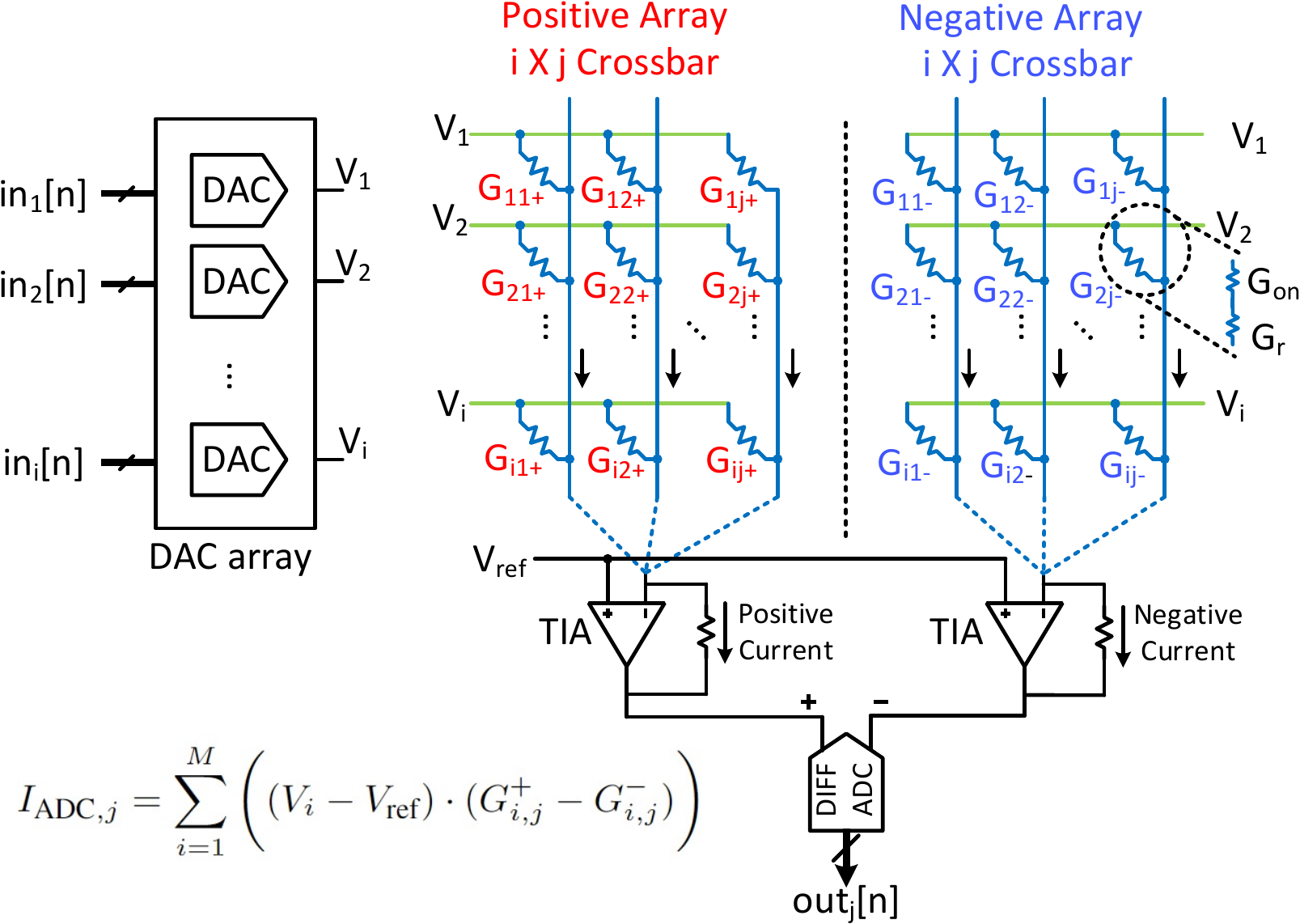}\vspace{-1.4em}
\caption{Analog IMC crossbar array pair (positive and negative arrays).}
\label{crossbar}\vspace{-2.5em}
\end{center}
\end{figure}

\textbf{Analog IMC Crossbar and Simulation Frameworks.} 
The resistive crossbar array (Fig.~\ref{crossbar}) forms the core computational primitive in IMC-based DNN accelerators, as it enables efficient parallel matrix-vector multiplications. In this architecture, DNN weights are encoded as the conductance of resistive cells, while activations are applied as voltages via DACs. Positive and negative weights are stored in separate arrays producing differential currents at the ADCs. The resulting MAC operation is expressed through $G_{i,j}^{\pm}$, denoting the conductance of cells in the arrays.  
Existing IMC simulation frameworks can be grouped into two categories (Table~\ref{comp}).  
$(i)$ \textit{System-level simulators}, such as NeuroSim \cite{chen2018neurosim}, DNN+NeuroSim V2.0 \cite{peng2020dnn+}, and MNSIM \cite{zhu2020mnsim}, use behavioral models to estimate area, power, and latency. MNSIM reports about 5\% deviation from SPICE in energy and latency for a two-layer network, while NeuroSim integrates with ML simulators to evaluate training and inference accuracy across memory types \cite{chen2018neurosim}. However, these tools lack analog non-ideality modeling and circuit-level fidelity. Similarly, IBM's AIHWKIT \cite{9458494}, though offering a PyTorch-based interface for crossbar simulation, inherits these limitations.  
$(ii)$ \textit{Circuit-level simulators}, such as DPE \cite{hu2016dot}, CCCS \cite{zhang2020cccs}, and IMAC-Sim \cite{Amin2023IMACSimAC}, provide higher accuracy by explicitly modeling device and circuit effects. DPE explores optimized weight mapping under non-idealities, while IMAC-Sim delivers a Python-based environment that generates SPICE netlists, incorporates parasitic interconnects, and supports partitioning, variability, and noise modeling. It enables both analog and digital emulation, with metrics such as power and accuracy extracted directly from HSPICE \cite{Amin2023IMACSimAC}. 

Despite progress in SPICE code generation~\cite{vungarala2024spicepilot,lai2024analogcoder}, IMC circuits pose unique challenges: even small networks require thousands of lines of repetitive SPICE code, and token limits restrict domain knowledge embedding. We address this by adopting IMAC-Sim's Python-based approach.

\noindent\textbf{Why LLMs Over Formal Methods.} While the IMC design space can be partially characterized by formal constraint satisfaction, two fundamental limitations motivate an LLM-based approach. First, \textit{scalability}: the full configuration space yields $|D_{\text{full}}| = 900{,}000$ configurations, requiring $\approx$13.7 years of exhaustive HSPICE simulation, rendering formal enumeration intractable. 
Second, \textit{soft constraint interpretation}:  user-specified intents such as ``edge deployment'' or  ``best accuracy under high variability'' are not formal predicates until semantically resolved. A formal solver requires constraints pre-specified in closed form; LIMCA infers design intent from natural language and maps it to valid parameter configurations autonomously. 
Critically, where formal and lookup-based methods exhibit a hard failure mode, returning null for unseen configurations. LIMCA's failure mode is soft: it proposes a candidate solution that is subject to iterative NHIL refinement, thereby providing a tractable starting point.%}

\vspace{-1.1em}

\section{LIMCA - The Proposed Framework}
\begin{figure}[t]\vspace{-1.8em}
  \centering
  \includegraphics[width=0.85\linewidth]    {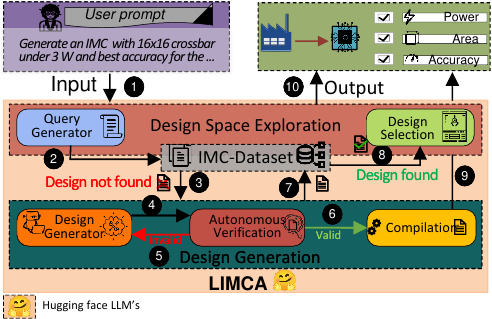}\vspace{-1.2em}
  \caption{\texttt{LIMCA} framework.}\vspace{-2.1 em}
  \label{LIMCA}
\end{figure}

\vspace{-0.4em}
\texttt{LIMCA} enables the automated design of analog IMC crossbars by leveraging LLMs to streamline design selection, generation, and verification, ensuring efficiency and adaptability in hardware-constrained environments. As illustrated in Fig.~\ref{LIMCA}, the framework enables both user-guided and autonomous design instantiating by dynamically interpreting user-defined constraints and optimizing IMC architectures accordingly.
The process begins with a user-specified prompt (\encircle{1}), defining key design requirements such as performance metrics, hardware constraints, and optimization goals. The LLM extracts relevant parameters and formulates a weighted query (\encircle{2}), determining whether an existing design from the design repository satisfies the given constraints. If an appropriate design is available, the system ranks and selects the optimal configuration (\encircle{8}), presenting it to the user (\encircle{10}). If no suitable design exists or if the user requests a new configuration, the framework triggers the Design Generation process (\encircle{3}). The LLM {selects and instantiates a novel IMC design configuration} that aligns with the specified constraints, generating a corresponding Python-based design representation.
The generated design undergoes an Automated Verification phase (\encircle{4}), where a script-driven NHIL validation assesses its correctness. If the design meets the required specifications (\encircle{6}), it is integrated into the design space repository (\encircle{7}), ensuring continuous expansion of the available solution space. In cases where verification fails, diagnostic feedback is generated, pinpointing errors and guiding the LLM in refining subsequent iterations, thereby reducing hallucinations and redundant modifications (\encircle{5}).
Unlike traditional design methodologies that rely on predefined architectures or manual fine-tuning, \texttt{LIMCA} dynamically adapts to evolving constraints, autonomously optimizing IMC designs while minimizing human intervention. This iterative and adaptive approach significantly enhances design efficiency, supporting various configurations and enabling scalable, high-performance IMC solutions. \texttt{LIMCA} is LLM-agnostic and can run with models available on Hugging Face~\cite{hugging_face}.

\begin{figure}[t] \vspace{-1.2em}
    \centering
    \includegraphics[width=0.95\linewidth]{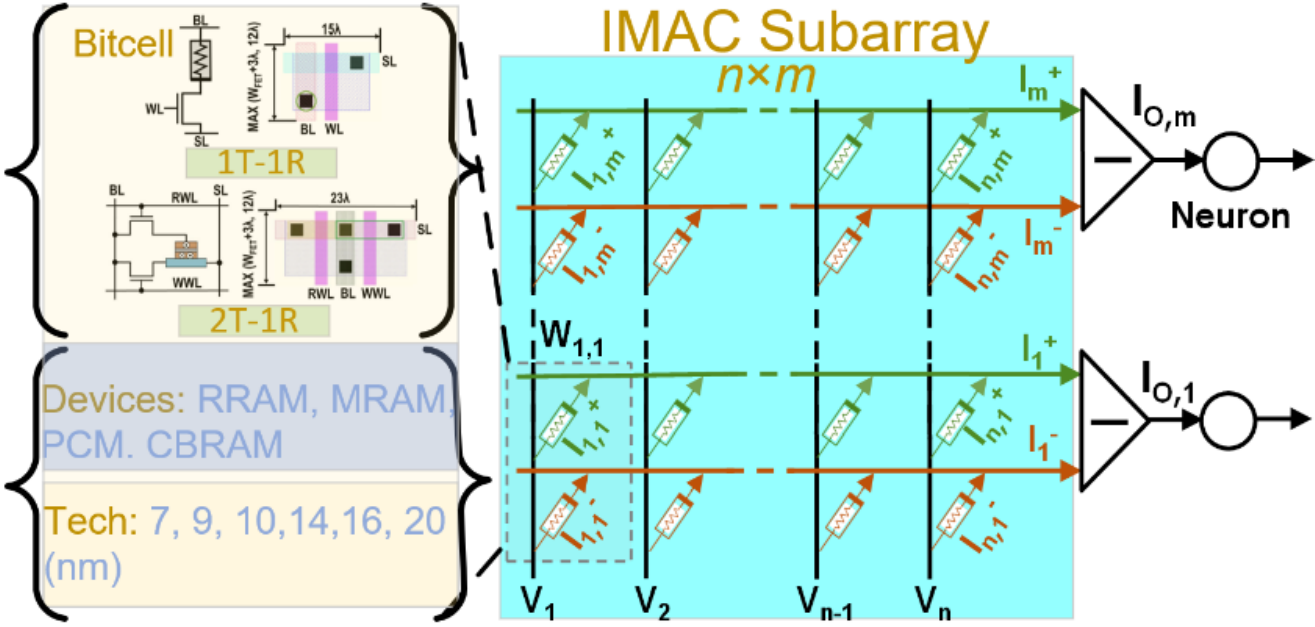}\vspace{-1em} 
    \caption{Overview of parameters across bitcells, devices, and technologies for IMC dataset curation.}\vspace{-2.4em} 
    \label{fig:ds}
\end{figure}

\textbf{IMC-Dataset.} For \texttt{LIMCA} to achieve user-constrained outputs, we construct a dedicated dataset, the \textit{IMC-Dataset}, to support the LLM. This dataset serves as a crucial component for integrating hardware-aware constraints into LLMs through either inference or fine-tuning. The dataset is built on IMAC-SIM~\cite{Amin2023IMACSimAC}, which facilitates both full analog circuit simulation and digital component emulation. It operates on the HSPICE Compiler, ensuring the generation of precise values as outlined in Table~\ref{comp}. To systematically explore the design space, we developed an automated framework to sweep key hardware parameters, capturing diverse configurations of IMC architectures. These parameters as shown in Fig.~\ref{fig:ds} includes different non-volatile memory devices (MRAM, RRAM, PCM, CBRAM), bit-cell configurations (1T-1R, 2T-1R, 1TG-1R), technology nodes, and bit-resolution(s). The dataset is primarily categorized based on crossbar size, with three distinct ($n \times m$) sizes. For each crossbar configuration, the following variations are considered: $3 \times 4 \times 3 \times 6=216$, resulting in $3 \times 216 = 648$ unique IMC instances across all three crossbar sizes. To extend the applicability of IMAC-SIM, we incorporate both digital and analog IMC variations. While the framework was initially designed for full analog simulations, we introduce an additional 216 analog data points by excluding the bit-resolution parameter, given the computational complexity of obtaining precise HSPICE simulation metrics across different bit-resolution(s). This ensures a broader exploration space while maintaining computational efficiency.

% \noindent To the best of our knowledge, this is the first open-source IMC dataset containing 400 analog and digital IMC variants. The dataset is generated from a single MLP topology; as the MLP architecture changes, the associated metrics adjust, enabling scalable extension. Planned expansions include additional MLP architectures and metrics derived from other classification tasks, broadening design-space coverage and evaluation. 
To the best of our knowledge, this is the first open-source IMC dataset with 400+ analog and digital variants. The dataset fixes the MLP configuration ($400 \times 120 \times 84 \times 10$) and $20 \times 20$ input dimensions for computational feasibility, though the design-generation pipeline adapts to different user requirements. Planned expansions include additional MLP architectures and classification tasks.
\noindent Because the combination of input feature-map size and MLP architecture exponentially increases the space (and compute), this release focuses on hardware exploration and avoids the high-dimensional complexity of varying network topologies across datasets. During creation, we fix the MLP configuration ($400 \times 120 \times 84 \times 10$), the dataset, and the number of MNIST training images~\cite{deng2012mnist}, ensuring consistent evaluation and computational feasibility. The design-generation and automated-validation pipeline itself is not restricted to these settings and can adapt to user requirements. 
\noindent Dataset metrics are strongly influenced by input image dimensions; here they are $20 \times 20$, given the specified MLP topology. 
% While conventional MLPs often use 784 inputs, we explore alternative configurations to optimize dataset generation via efficient horizontal and vertical crossbar partitioning.

{\textbf{LLM Adaptation and Error Mitigation.} Our fine-tuning-free approach intentionally demonstrates that effective IMC design automation is achievable with off-the-shelf LLMs, reducing the barrier to adoption and enabling rapid deployment without expensive model customization. This aligns with recent findings that in-context learning can match fine-tuned performance on domain-specific tasks when provided with well-structured exemplars~\cite{dai2023icl}. Our LLM adaptation strategy employs: $(i)$ \textit{Structured Prompting}---we decompose the design task into modular sub-tasks (parameter extraction $\rightarrow$ constraint interpretation $\rightarrow$ code generation $\rightarrow$ verification), reducing cognitive load and error propagation at each step; $(ii)$ \textit{In-Context Learning}---we provide the LLM with exemplar (X, Y) pairs from the dataset sorted by relevance to the query using k-nearest neighbor retrieval~\cite{izacard2023atlas}, enabling pattern recognition without gradient updates; $(iii)$ \textit{NHIL Error Correction}---our pipeline automatically validates generated code against syntax correctness, parameter range validity, and constraint satisfaction, with failed generations triggering diagnostic feedback specifying error type and location; %\rev{
The effectiveness of NHIL correction is directly quantified by the Pass@1 versus Pass@3 performance gap reported in Tables~\ref{LIMCA_table_exploration} and~\ref{tab:baseline}: Pass@1 reflects single-shot generation without correction, while Pass@3 captures iterative NHIL refinement. The consistent convergence to 100\% Pass@3 across all models demonstrates that the feedback loop deterministically resolves the dominant failure categories, syntax errors (38\%) and parameter range violations (31\%), which together constitute 69\% of all generation failures, eliminating the need for human intervention in error diagnosis and correction.%}
and $(iv)$ \textit{Hallucination Mitigation}---by constraining the LLM to generate Python code calling IMAC-Sim APIs rather than raw SPICE netlists, we bound the output space to valid configurations, preventing hallucination of non-existent device types or invalid parameter combinations.} \vspace{-1.2em}

\begin{table*}[t] \vspace{-2.8em}
\caption{Design space explored by LIMCA on MNIST; entries meeting power $\leq$3W and accuracy $\geq$96\% are highlighted.\vspace{-0.9em}}
\centering
\scalebox{0.72}{
\rowcolors{2}{gray!10}{white}
\begin{tabular}{ccc|ccc|ccc|ccc}
\toprule
\multicolumn{3}{c|}{Config./Xbar Size} & \multicolumn{3}{c|}{16$\times$16} & \multicolumn{3}{c|}{32$\times$32} & \multicolumn{3}{c}{64$\times$64} \\ 
\midrule
Tech & Device & Bitcell & Area ($\mu$m$^2$) & Accuracy (\%) & Power (W) & Area ($\mu$m$^2$) & Accuracy (\%) & Power (W) & Area ($\mu$m$^2$) & Accuracy (\%) & Power (W) \\ 
\midrule
7nm & MRAM & 1T1R & 5286.615 & \AccCell{96} & \PowCell{3.937868} & 3006.403 & \AccCell{96} & \PowCell{3.101278} & 2156.134 & \AccCell{82} & \PowCell{1.847222} \\
7nm & RRAM & 1T1R & 5286.615 & \AccCell{78} & \PowCell{8.291856} & 3006.403 & \AccCell{62} & \PowCell{5.490012} & 2156.134 & \AccCell{18} & \PowCell{2.915078} \\
7nm & RRAM & 2T1R & 5602.122 & \AccCell{80} & \PowCell{8.161842} & 3329.135 & \AccCell{52} & \PowCell{5.458412} & 2541.486 & \AccCell{14} & \PowCell{2.18464} \\
7nm & PCM & 1T1R & 5286.615 & \AccCell{92} & \PowCell{0.53445} & 3006.403 & \AccCell{98} & \PowCell{0.521569} & 2156.134 & \AccCell{100} & \PowCell{0.457961} \\
7nm & PCM & 2T1R & 5602.122 & \AccCell{92} & \PowCell{0.533303} & 3329.135 & \AccCell{98} & \PowCell{0.521374} & 2541.486 & \AccCell{100} & \PowCell{0.778821} \\ 
\midrule
9nm & MRAM & 1T1R & 5672.95 & \AccCell{94} & \PowCell{4.041462} & 3401.585 & \AccCell{96} & \PowCell{3.250092} & 3265.004 & \AccCell{72} & \PowCell{1.987902} \\
9nm & RRAM & 1T1R & 5672.95 & \AccCell{100} & \PowCell{8.618146} & 3401.585 & \AccCell{68} & \PowCell{5.894228} & 2627.994 & \AccCell{18} & \PowCell{3.171676} \\
9nm & RRAM & 2T1R & 6194.502 & \AccCell{86} & \PowCell{7.372028} & 3935.08 & \AccCell{62} & \PowCell{7.89253} & 3265.004 & \AccCell{14} & \PowCell{3.153108} \\
9nm & PCM & 1T1R & 5672.95 & \AccCell{98} & \PowCell{0.535587} & 3401.585 & \AccCell{98} & \PowCell{0.525815} & 2627.994 & \AccCell{100} & \PowCell{0.469902} \\
9nm & PCM & 2T1R & 6194.502 & \AccCell{82} & \PowCell{0.533361} & 3935.08 & \AccCell{98} & \PowCell{0.525645} & 3265.004 & \AccCell{100} & \PowCell{0.469741} \\ 
\midrule
14nm & MRAM & 1T1R & 7061.34 & \AccCell{98} & \PowCell{4.087762} & 4821.77 & \AccCell{96} & \PowCell{3.464416} & 4323.738 & \AccCell{96} & \PowCell{2.228876} \\
14nm & RRAM & 1T1R & 7061.34 & \AccCell{86} & \PowCell{9.062472} & 4821.77 & \AccCell{84} & \PowCell{6.528764} & 4323.738 & \AccCell{24} & \PowCell{3.606136} \\
14nm & RRAM & 2T1R & 8323.367 & \AccCell{84} & \PowCell{4.244777} & 6112.698 & \AccCell{64} & \PowCell{6.498698} & 5865.144 & \AccCell{18} & \PowCell{3.587574} \\
14nm & PCM & 1T1R & 7061.34 & \AccCell{90} & \PowCell{0.536243} & 4821.77 & \AccCell{98} & \PowCell{0.531266} & 4323.738 & \AccCell{100} & \PowCell{0.486095} \\
14nm & PCM & 2T1R & 8323.367 & \AccCell{94} & \PowCell{0.542201} & 6112.698 & \AccCell{98} & \PowCell{0.53113} & 5865.144 & \AccCell{100} & \PowCell{0.485948} \\ 
\midrule
20nm & MRAM & 1T1R & 9524.224 & \AccCell{98} & \PowCell{4.148678} & 7341.056 & \AccCell{96} & \PowCell{3.596336} & 7331.84 & \AccCell{96} & \PowCell{2.39336} \\
20nm & RRAM & 1T1R & 9524.224 & \AccCell{92} & \PowCell{3.304907} & 7341.056 & \AccCell{88} & \PowCell{6.954812} & 7331.84 & \AccCell{46} & \PowCell{3.924754} \\
20nm & RRAM & 2T1R & 12099.79 & \AccCell{90} & \PowCell{2.468912} & 9975.603 & \AccCell{70} & \PowCell{6.787644} & 10477.57 & \AccCell{22} & \PowCell{3.906236} \\
20nm & PCM & 1T1R & 9524.224 & \AccCell{96} & \PowCell{0.537193} & 7341.056 & \AccCell{98} & \PowCell{0.534285} & 7331.84 & \AccCell{100} & \PowCell{0.495511} \\
20nm & PCM & 2T1R & 12099.79 & \AccCell{98} & \PowCell{0.538646} & 9975.603 & \AccCell{98} & \PowCell{0.534169} & 10477.57 & \AccCell{100} & \PowCell{0.495376} \\ 
\bottomrule
\end{tabular}}
\label{exp}\vspace{-2.5em}
\end{table*}

\section{Experimental Analysis}
\vspace{-0.5em}

\begin{figure}[t]
  \centering
  \includegraphics[width=0.85\linewidth]{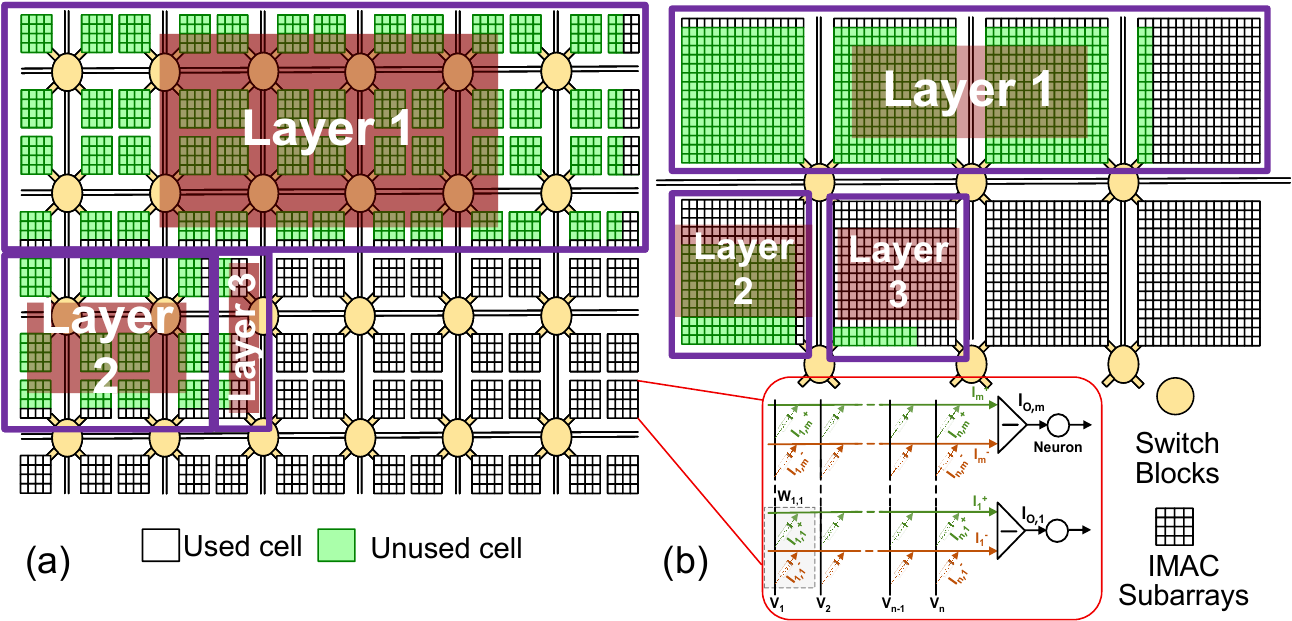}\vspace{-1.2em}
  \caption{Deployment of a 400$\times$120$\times$84$\times$10 MLP on fully-analog IMC architecture with (a) 32$\times$32, (b) 128$\times$128 subarrays.}\vspace{-2em}
  \label{partition}
\end{figure}

\textbf{Design Space Exploration.}\label{4-a}
The queries driving the DSE are generated by ChatGPT-4o \cite{CHATGPT}. A total of 30 queries are systematically divided based on specific objectives: 10 queries prioritize power efficiency with relaxed technology node constraints, 10 emphasize area optimization, and 10 enforce hard constraints on design feasibility. The selection schema of the DSE for IMC accelerators, inferring from the dataset by \texttt{LIMCA}, is summarized in Table \ref{exp}. From Table~\ref{exp}, we observe area reduction with increasing crossbar size due to improved partitioning that reduces switch-box overhead (Fig.~\ref{partition}). Considering edge vision sensor requirements, a nominal power limit of $\leq$3W is imposed while targeting accuracy $\geq$96\%. Entries in green satisfy these constraints; yellow highlights near-optimal configurations (e.g., 64$\times$64 crossbar with 1T1R-PCM at 7nm or 9nm). Table~\ref{LIMCA_table_exploration} summarizes \texttt{LIMCA}'s DSE performance across multiple LLMs. All achieve 10/10 \textit{Pass@3} across constraint categories, indicating reliable design selection within 3 attempts. \textit{Pass@1} varies: \textit{Qwen2.5-Coder-32B-Instruct} leads with $\geq$9/10 across categories, while \textit{Mamba-Codestral-7B-v0.1} shows consistent 8/10 performance.

\begin{table}[b] \vspace{-1.8em}
\centering
\caption{LIMCA Performance Results - DSE.} \vspace{-1em}
\label{LIMCA_table_exploration}
\scalebox{0.75}{
\begin{tabular}{llccc}
\hline
\multirow{2}{*}{\textbf{Model}} & \multirow{2}{*}{\textbf{Metric}} & \multicolumn{3}{c}{\textbf{Query Focus}} \\
\cline{3-5}
& & Power & Area & Hard Constraints \\
\hline
\textbf{Qwen2.5-7B-Instruct-1M} & Pass@1 & 9/10 & 9/10 & 8/10 \\
& Pass@3 & 10/10 & 10/10 & 10/10 \\
\hline
\textbf{DeepSeek-R1-Distill-Qwen-1.5B} & Pass@1 & 8.3/10 & 9/10 & 7.9/10 \\
& Pass@3 & 10/10 & 10/10 & 10/10 \\
\hline
\textbf{Mamba-Codestral-7B-v0.1} & Pass@1 & 8/10 & 8/10 & 8/10 \\
& Pass@3 & 10/10 & 10/10 & 10/10 \\
\hline
\textbf{Qwen2.5-Coder-32B-Instruct} & Pass@1 & 9/10 & 9/10 & 9.3/10 \\
& Pass@3 & 10/10 & 10/10 & 10/10 \\
\hline
\textbf{Llama-3.1-8B-Instruct} & Pass@1 & 9/10 & 8.6/10 & 8.6/10 \\
& Pass@3 & 10/10 & 10/10 & 10/10 \\
\hline
\end{tabular}}\vspace{-2.7em}
\end{table}

{\textbf{Baseline Comparison.} We evaluate \texttt{LIMCA} against three DSE baselines: exhaustive Grid Search, Genetic Algorithm (GA), and a Rule-Based system. The GA employs a population of 50, tournament selection ($k=3$), single-point crossover (0.8), bit-flip mutation (0.1), and terminates at 100 generations or upon 5-generation stagnation. Soft constraints (e.g., "edge deployment" $\rightarrow$ power $<$ 3W) are managed via a weighted fitness function with quadratic penalties. Under an equivalent budget of $\sim$500 evaluations, Grid Search and GA are limited to existing database configurations, failing on extrapolation queries (0\% success) and struggling with soft constraints (45\% and 62\%, respectively). In contrast, \texttt{LIMCA} leverages natural language understanding to achieve 93\% on soft constraints and 100\% on extrapolation, autonomously generating novel, valid designs where traditional methods return "not found."}

\begin{table}[h] \vspace{-2em}
\centering
{\caption{Baseline Comparison on Query Types.}} \vspace{-1em}
\label{tab:baseline}
{\scalebox{0.82}{
\begin{tabular}{lccc}
\hline
\textbf{Method} & \textbf{Exact Match} & \textbf{Soft Constraints} & \textbf{Extrapolation}$^\dagger$ \\
\hline
Grid Search & 100\% & 45\% & 0\% \\
Genetic Algorithm & 98\% & 62\% & 0\% \\
Rule-Based & 95\% & 67\% & 0\% \\
\textbf{LIMCA} & \textbf{100\%} & \textbf{93\%} & \textbf{100\%} \\
\hline
\end{tabular}}\\[0.3em]
\footnotesize{$^\dagger$Percentage of queries receiving a proposed solution. Traditional methods return ``not found'' for configurations outside the database.}}
\vspace{-1em}
\end{table}

\textbf{Design Creation.}
\texttt{LIMCA} synthesizes IMC designs with automated validation. Design requests originate from ChatGPT queries and are executed by \texttt{LIMCA}, which verifies each result against specified constraints. Table~\ref{tab:limca_DG} reports \textit{Pass@1} and \textit{Pass@3} for design generation and verification. For generation, \texttt{LIMCA} attains $\geq$89\% \textit{Pass@1} and 100\% \textit{Pass@3}, indicating reliable synthesis. For verification, we inject erroneous designs to stress-test the framework; \texttt{LIMCA} detects and corrects errors with 91.5\% average \textit{Pass@1} and 100\% \textit{Pass@3}, demonstrating robust NHIL-based error recovery. {Failure analysis of 500 generation attempts reveals four primary modes: syntax errors (38\%), parameter range violations (31\%), device type hallucinations (18\%), and constraint mismatches (13\%), with the first two categories (69\%) being deterministically correctable given NHIL diagnostic feedback.}

{\textbf{Simulation Methodology.} Power and area metrics are derived from HSPICE simulation of the complete crossbar netlist. Inference accuracy is computed via behavioral simulation using SPICE-characterized device models applied to a representative subset of test images---not a full transistor-level simulation of 10,000 images, which would be computationally infeasible. This  approach provides SPICE-level fidelity for circuit metrics while enabling practical accuracy evaluation.}

\begin{table}[t] \vspace{-1.0em}
\centering
\caption{Design Generation and Automated Verification.} \vspace{-0.5em}
\label{tab:limca_DG}
\scalebox{0.8}{
\begin{tabular}{lcc|cc}
\hline
\multirow{2}{*}{\textbf{Model}} & \multicolumn{2}{c|}{\textit{\textbf{Design Generation}}} & \multicolumn{2}{c}{\textit{\textbf{Design Verification}}} \\ 
\cline{2-5} 
& \textbf{Pass@1} & \textbf{Pass@3} & \textbf{Pass@1} & \textbf{Pass@3} \\ 
\hline
\textbf{Qwen2.5-7B-Instruct-1M}        & 89\%   & 100\%  & 90\%   & 100\%  \\ 
\textbf{DeepSeek-R1-Distill-Qwen-1.5B} & 92\%   & 100\%  & 94\%   & 100\%  \\ 
\textbf{Mamba-Codestral-7B-v0.1}       & 91\%   & 100\%  & 89\%   & 100\%  \\ 
\textbf{Qwen2.5-Coder-32B-Instruct}    & 95\%   & 100\%  & 91\%   & 100\%  \\ 
\textbf{Llama-3.1-8B-Instruct}         & 96\%   & 100\%  & 94\%   & 100\%  \\ 
\hline
\end{tabular}%
} \vspace{-1.5em}
\end{table}

{\textbf{Scalability Analysis.}  The IMC design space is characterized by the configuration vector $\mathbf{X}$ = (\text{crossbar size}, \text{tech node}, \text{bitcell}, \text{device}, \text{MLP topology}, \text{pixel dimensions}, \text{dataset}). While our current dataset fixes the MLP topology and pixel dimensions for computational feasibility, the full design space is combinatorially large. With realistic parameter ranges $|X_{\text{crossbar}}| = 6$ sizes (16$\times$16 through 256$\times$256), $|X_{\text{tech}}| = 6$ nodes (7nm--45nm), $|X_{\text{bitcell}}| = 4$ configurations, $|X_{\text{device}}| = 5$ types, $|X_{\text{MLP}}| \geq 50$ topologies, $|X_{\text{pixel}}| = 5$ dimensions, $|X_{\text{dataset}}| = 5$ benchmarks---the full design space yields $|D_{\text{full}}| = 900{,}000$ configurations.  At approximately 8 minutes per HSPICE simulation, exhaustive enumeration would require $\approx$13.7 years of continuous computation. This intractability motivates \texttt{LIMCA}'s approach: rather than enumerating all configurations, the LLM learns to navigate the design space intelligently, proposing valid solutions for queries and self-correcting through NHIL feedback.}

{\textbf{Generalization Validation.} To demonstrate that \texttt{LIMCA} generalizes beyond memorization, we conducted hold-out experiments removing entire configuration categories (e.g., all 64$\times$64 crossbars, all PCM devices) from the LLM's context. Unlike Grid Search, which returns ``not found'' for any configuration outside its database, \texttt{LIMCA} proposes solutions for 100\% of such queries. While extrapolation to fundamentally different device physics remains challenging (e.g., PCM's 0.5W power profile versus RRAM's 3--8W), \texttt{LIMCA} demonstrates genuine pattern recognition for interpolation within the known design space, autonomously navigating toward constraint-satisfying configs via iterative NHIL refinement.}

{\textbf{Trade-off Reasoning.} Beyond selection, \texttt{LIMCA} articulates device trade-offs. For the query ``best accuracy under high variability,'' \texttt{LIMCA} responds: ``PCM achieves 98--100\% accuracy across crossbar sizes due to tighter conductance distribution, while RRAM accuracy drops from 100\% at 16$\times$16 to 18\% at 64$\times$64. \textbf{Recommendation:} PCM with 1T1R at 7nm (0.46W, 100\% accuracy).'' This demonstrates the LLM learning/ articulating relationships between device physics and performance.}

\textbf{Design Exploration Cost.}
To assess DSE time under identical constraints ($\leq$3W power, $\geq$96\% accuracy), a circuit expert conducted manual optimization using CCSS \cite{zhang2020cccs} and IMAC-Sim \cite{Amin2023IMACSimAC}. Each iteration required debugging, reconfiguration, and reruns, with total optimization ranging from 140--398 minutes (CCSS) and 92--154 minutes (IMAC-Sim). \texttt{LIMCA} retrieves existing configurations in seconds or synthesizes new designs in under 8 minutes (Table~\ref{dse}), achieving 11.5$\times$--49.7$\times$ speedup while eliminating manual intervention.

\begin{table}[t]\vspace{-0.4em}
\centering
\caption{Estimated DSE Time.} \vspace{-0.8em}
\scalebox{0.8}{
\begin{tabular}{lccc}
\hline
\multicolumn{1}{c}{Frameworks} & \begin{tabular}[c]{@{}c@{}}Design Space\\ Exploration\end{tabular} & Language & \begin{tabular}[c]{@{}c@{}}Experiment\\ Time (minutes)\end{tabular} \\ \hline
CCSS \cite{zhang2020cccs} & Manual & \multicolumn{1}{l}{MATLAB-SPICE} & $140\leq t_{DSE}\leq398$ \\
IMAC-Sim \cite{Amin2023IMACSimAC} & Manual & Python-SPICE & $92\leq t_{DSE}\leq154$ \\
\texttt{LIMCA} & Automated & Python-SPICE & $t_k\leq t_{DSE}\leq8$ \\ \hline
\end{tabular}}
\label{dse}\vspace{-2em}
\end{table}

\vspace{-1em}

\section{Conclusion}\vspace{-0.4em}
In conclusion, this work introduces \texttt{LIMCA}, a novel open-sourced framework that leverages LLMs to automate the design and evaluation of IMC crossbar architectures with no human intervention in design synthesis and SPICE-level verification. By systematically generating and validating SPICE-characterized configurations, \texttt{LIMCA} accelerates design exploration while satisfying key power, area, and accuracy constraints. Baseline comparisons demonstrate significant advantages over grid search and rule-based approaches.
% while traditional methods return ``not found'' for configurations outside their databases, \texttt{LIMCA} proposes solutions for 100\% of queries and autonomously navigates toward valid designs through NHIL feedback.
\vspace{-1em}

\section{Acknowledgments}%\vspace{-0.5em}
\small The authors used Claude Sonnet only to improve writing clarity and readability, while all scientific content, including ideas, experiments, and conclusions, was created entirely by the authors. \vspace{-1 em}
% \begin{table}[t]
% \centering
% \caption{Estimated DSE Time.} \vspace{-0.5em}
% \scalebox{0.8}{
% \begin{tabular}{lccc}
% \hline
% \multicolumn{1}{c}{Frameworks} & \begin{tabular}[c]{@{}c@{}}Design Space\\ Exploration\end{tabular} & Language & \begin{tabular}[c]{@{}c@{}}Experiment\\ Time (minutes)\end{tabular} \\ \hline
% CCSS \cite{zhang2020cccs} & Manual & \multicolumn{1}{l}{MATLAB-SPICE} & $140\leq t_{DSE}\leq398$ \\
% IMAC-Sim \cite{Amin2023IMACSimAC} & Manual & Python-SPICE & $92\leq t_{DSE}\leq154$ \\
% \texttt{LIMCA} & Automated & Python-SPICE & $t_k\leq t_{DSE}\leq8$ \\ \hline
% \end{tabular}} \sc
% \label{dse}\vspace{-2em}
% \end{table}

\bibliographystyle{IEEEtran}
\bibliography{references}

\end{document}